\documentclass[]{spie}  

\usepackage{amsmath,amsfonts,amssymb}
\usepackage{graphicx}
\usepackage[colorlinks=true, allcolors=blue]{hyperref}
\usepackage{enumitem}

\title{Inverse mask design for interference lithography using automatic differentiable wave propagation}

\author[a]{Chuntian Cao}
\author[a]{Jangwoon Sung}
\author[a]{Jack Griffiths}
\author[a]{Yuan Gao}
\author[a]{Xi Yu}
\author[a]{Paul Baity}
\author[a]{Nikhil Tiwale}
\author[a]{Zhitian Shi}
\author[a]{Juhong Ahn}
\author[a]{Shinjae Yoo}
\author[a]{Yong S. Chu}
\author[a]{Chang-Yong Nam}
\affil[a]{Brookhaven National Laboratory, Upton, New York 11973, United States}

\authorinfo{Further author information: (Send correspondence to C.C.)\\C.C.: E-mail: ccao@bnl.gov}

\begin{document} 
\maketitle

\begin{abstract}

Interference lithography (IL) is powerful for fabricating high-resolution periodic nanostructures, but designing masks to produce non-periodic patterns remains challenging. 
We introduce a gradient-based optimization framework for binary IL mask design using automatic differentiation. 
The forward model is implemented using the differentiable angular spectrum method (ASM). 
The inverse mask design is formulated as an optimization problem, where the mask logits are updated through backpropagation of the loss between the simulated field amplitude and the target pattern. 
We optimize a mask that reproduces a target pattern with only 0.1\% isolated pixel-level defects, resolving features at half the mask pixel pitch. 
To scale mask optimization, we employ the shifted ASM, which partitions the mask into patches that are propagated independently and summed at the image plane. 
For a $3.84~\text{mm} \times 3.84~\text{mm}$ mask, shifted ASM with 16 patches reduces peak GPU memory by $3.8\times$ at only $1.3\times$ runtime cost relative to standard ASM. With gradient checkpointing, peak memory is reduced by $7.4\times$ at $2\times$ runtime. 
Distributing across multiple GPUs further accelerates the optimization. 
This work establishes a physics-informed, machine learning-driven approach for IL mask design, moving a step further towards complex, non-periodic patterns. The source code is available at 
\url{https://github.com/chuntian236/holography-optimization.git}.

\end{abstract}


\section{INTRODUCTION}
\label{sec:intro}  

Interference lithography (IL) creates high-resolution patterns by interfering two or more coherent beams on a photosensitive layer without the projection optics required by conventional lithography~\cite{MOJARAD201555, Ekinci_Nanoscale_2024}.
IL has become a standard tool for producing periodic nanostructures and for developing and qualifying photoresists at extreme ultraviolet (EUV) wavelengths. 
Extending IL to arbitrary, non-periodic patterns has recently become a focus for extending the flexibility and reliability of nano-scale lithographic designs~\cite{li2026holographiceuvlithography40, Cao_NYSDS}. While the forward propagation from a photomask to aerial image can be computed analytically, the inverse mask design for a target aerial image is an ill-posed problem, due to the lack of phase information in the aerial patterns. 

In this work, we solve the mask design problem with gradient-based optimization through a differentiable forward propagation model. 
We use the angular spectrum method (ASM) as the forward propagator.  
The ASM is a standard technique for simulating the near- and intermediate-field diffraction relevant to IL~\cite{Matsushima:09, Matsushima:10}. 
We optimize the mask by backpropagating the difference between the simulated field amplitude and the target pattern, implemented within an automatic-differentiation framework.
This gradient-based, physics-informed optimization strategy has been applied for phase retrieval and ptychographic reconstruction~\cite{Wu:24, yao2022autophasenn, 10.1117/12.3051223}, and for computational photomask and optical-proximity-correction design in projection lithography~\cite{chen2024opensource}. 
 
With this approach, we optimize a mask that reproduces a target pattern with isolated pixel-level defects in only 0.1\% of the pattern (Section~\ref{sec:res_asm}). To simulate fields larger than a single Fourier transform can hold in memory, we use the shifted ASM~\cite{Muffoletto:07, Matsushima:10} for large mask sizes. For a moderate mask (3.84~mm~$\times$~3.84~mm), this reduces peak GPU memory by $3.8\times$ at a $1.3\times$ runtime cost, or by $7.4\times$ with gradient checkpointing at $2\times$ runtime, relative to standard ASM. 
Using 16 patches, we optimize masks as large as 12.8~mm~$\times$~12.8~mm, a $10.2\times$ larger area than the largest mask that fits in memory with standard ASM. Distributing the same optimization across 4 GPUs further accelerates the optimization, making the optimization $2.5\times$ faster relative to standard ASM.

\section{Approach}

\subsection{Overview}
\label{sec:overview-approach}

\begin{figure} [ht]
   \begin{center}
   \begin{tabular}{c} 
   \includegraphics[height=9cm]{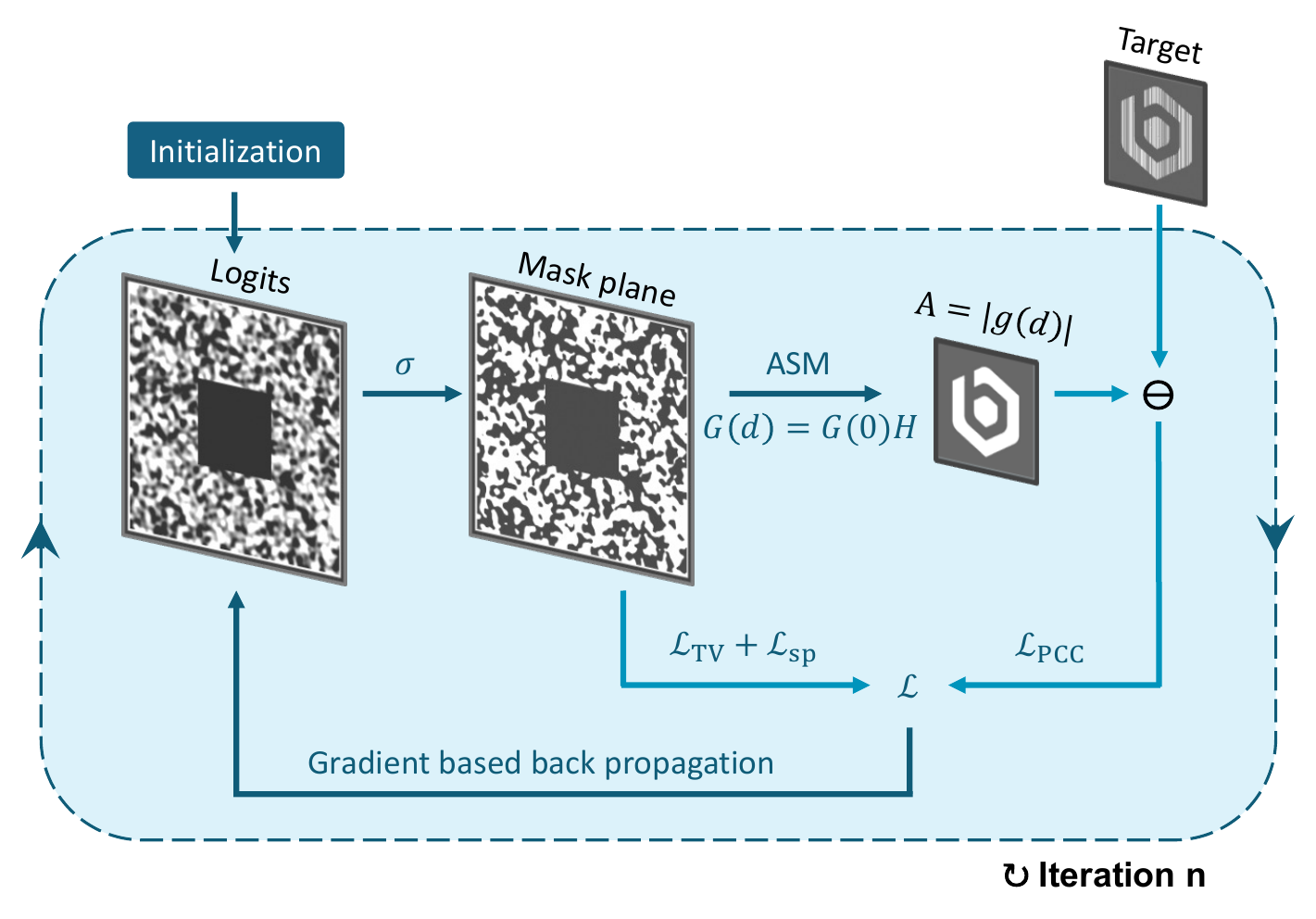}
   \end{tabular}
   \end{center}
   \caption[example] 
   { \label{fig:workflow} 
Schematic of the gradient-based inverse mask design pipeline. 
The mask logits are initialized from a spatially correlated Gaussian random field. 
At each iteration, the logits are converted to a near-binary mask via a sigmoid function $\sigma$, and the wavefront at the image plane $g(d)$ is computed by propagating the mask field using the angular spectrum method (ASM). 
The loss function comprises the Pearson correlation loss $\mathcal{L}_{PCC}$ between the simulated field amplitude $A=|g(d)|$ and the target pattern, the total variation loss $\mathcal{L}_{TV}$, and the sparsity loss $\mathcal{L}_{sp}$. 
$\mathcal{L}_{TV}$ and $\mathcal{L}_{sp}$ are regularization losses acting on the mask. 
The loss is backpropagated through the differentiable forward model to update the logits. }
   \end{figure}

Figure~\ref{fig:workflow} shows an overview of the workflow. 
We use automatic differentiation approach to design a mask that can produce the target pattern. 
The approach includes a forward simulation to calculate the propagated field from the mask field, 
and an inverse optimization to design the mask from target pattern. 
In each iteration, the mask is updated through gradient-based learning, using gradients computed from the loss between the simulated field amplitude and the target pattern. 
The forward simulation is introduced in \ref{sec:ASM} and \ref{sec:shifted-ASM}, and the inverse optimization is introduced in \ref{sec:optimization}; relevant equations are given in Appendix~\ref{app:propagation} (propagation models) and Appendix~\ref{app:loss} (loss function). 

\subsection{Angular Spectrum Method (ASM)}
\label{sec:ASM}

The ASM for forward simulation is briefly summarized here, largely following the notations in Matsushima's work~\cite{Matsushima:10}. The full formulation, including the anti-aliasing conditions, is given in Appendix~\ref{app:ASM}.
The present study uses a scalar-field approximation. At high numerical aperture (NA), vectorial diffraction and polarization-dependent mask responses may become non-negligible and are not included in the current model.

Given the complex field amplitude $g(x,y;z=0)$ at the mask plane, ASM computes the field $g(x,y;z=d)$ at the image plane in three steps: (1) Fourier transform $g$ to obtain its angular spectrum $G(u,v;0)$; 
(2) multiply $G$ by a transfer function $H(u,v;d)$ that combines free-space propagation with a band-limiting window suppressing evanescent and aliased frequency components; 
and (3) inverse Fourier transform the result to recover the propagated field $g(x,y;z=d)$. 
The ASM method is based on differentiable formulas, making gradient-based optimization possible. 

\subsection{Shifted ASM for Memory-Efficient Propagation}
\label{sec:shifted-ASM}

One shortcoming of the ASM is its memory requirement. 
Since the Fourier transform has to be performed over the entire image, the working memory scales as $\mathcal{O}(N^2)$ for a mask of size $N \times N$ after padding, which becomes prohibitive for large masks.

One way to resolve this is to use the shifted ASM~\cite{Muffoletto:07, Matsushima:10} as the propagator. 
The mask is partitioned into $P \times P$ patches. Each patch is propagated independently on a grid of size $(N/P)^2$ using a transfer function $\hat{H}(u, v;\, d)$ that accounts for the patch's lateral offset from the observation window. 
The resulting complex fields are summed at the observation plane. 
Peak memory then scales with the patch size, $\mathcal{O}\big((N/P)^2\big)$. 
The full shifted transfer function and its anti-aliasing conditions are given in Appendix~\ref{app:shifted-ASM}.


\subsection{Gradient-Based Mask Optimization} 
\label{sec:optimization}

In this work, the incident illumination wavefront is set as a plane wave $g_0(x,y)=1$ for simplicity; $g_0$ can be substituted with other values, such as Gaussian beam illumination or a tilted plane wave, without changing the optimization procedure. The photomask transmittance $M(x,y) \in \{0,1\}$ modulates the illumination, so the field at the mask plane is $g(x,y;0) = M(x,y)$. 

The aerial image at the image plane is the intensity of the propagated field,
\begin{equation}
    I(x, y) = \left|g(x, y;\, d)\right|^2. 
\end{equation}
and the field amplitude is $A(x,y) = |g(x,y;\,d)|$.

Given a target pattern $I_t(x,y)$, the goal is to find a mask $M$ such that the produced field amplitude $A(x,y) \approx I_t(x,y)$.
For a binary target pattern $I_t$, the global optimum is the same whether $A$ or $I$ is optimized. 
However, the gradients and loss landscapes differ. 
We choose to optimize the field amplitude $A(x,y)$ in the loss function, because $I = A^2$ suppresses the differences in low intensity regions and reduces their relative weighting in the loss, whereas $A$ preserves the dynamic range more uniformly across the image. 


We use a gradient-based approach, optimizing over continuous logits $\ell(x,y)$ that are converted to a soft mask through a sigmoid function with temperature $\tau$:
\begin{equation}
    \tilde{M}(x,y) = \sigma\!\left(\ell(x,y) / \tau\right),
    \label{eq:mask_sigmoid}
\end{equation}
which approximates a binary mask as $\tau$ is annealed during optimization (Section~\ref{subsec:asm_opt}); the hard binary mask used at evaluation is $M = \mathbf{1}[\tilde{M} > 0.5]$.

The logits $\ell(x,y)$ are initialized as a spatially correlated Gaussian random field, 
generated by convolving Gaussian noise with a Gaussian kernel of width $\sigma_c$. 
The central region of the mask is fixed to be opaque throughout optimization, because light passing through the center of the mask contributes only low-spatial-frequency components, which do not contribute to high-resolution feature formation. 
As in many gradient-based optimization problems, the initialization condition could affect the final optimized result. 
For instance, initializing the mask with fewer ``on''pixels can add sparsity to the final mask, facilitating mask manufacturability (Section~\ref{sec:mas-degen}).

At each iteration, the simulated field amplitude $A$ (computed with ASM or shifted ASM) is compared with the binary target pattern $I_t$.
The loss $\mathcal{L}$ combines a Pearson-correlation term $\mathcal{L}_\mathrm{PCC}$ with regularization terms,  including total-variation $\mathcal{L}_\mathrm{TV}$ and sparsity $\mathcal{L}_\mathrm{sp}$, favoring manufacturable masks. 
The explicit form of each term are given in Appendix~\ref{app:loss}. The logits are updated by backpropagating $\mathcal{L}$ through the differentiable forward model using the Adam optimizer.~\cite{kingma2015adam}

\section{Results}

\subsection{Optimization Using ASM}
\label{sec:res_asm}

We show that gradient-based optimization of the mask logits, using ASM as the differentiable forward model, can find a binary mask that reproduces the target pattern at the image plane. 

\subsubsection{Simulation Parameters}

We use an illumination wavelength of $\lambda$= 445 nm, corresponding to blue laser. 
The model can be applied to lithography / holography by changing the wavelength. For example, using $\lambda$ = 13.5 nm corresponds to EUV lithography.
The Nyquist sampling criterion requires a pixel size smaller than $\lambda/2$. 
We use $\Delta x = \Delta y = 200$~nm  at the image plane and during propagation. 
The mask contains $19200\times 19200$ pixels, corresponding to a mask size of $D^2 = 3.84~\text{mm} \times 3.84~\text{mm}$. 
The propagation distance is $d = 2 $~mm. The numerical aperture is calculated to be: 
\begin{equation}
    \text{NA}=\sin(\arctan(\frac{D/2}{d}))=0.69.
\end{equation}
To suppress aliasing in the discrete Fourier transform, the field is zero-padded by 5760 pixels on each side (30\% of the field size) before propagation, resulting in $N=30720$ on each side for padded masks.

\begin{figure} [htbp]
   \begin{center}
   \includegraphics[width=\textwidth]{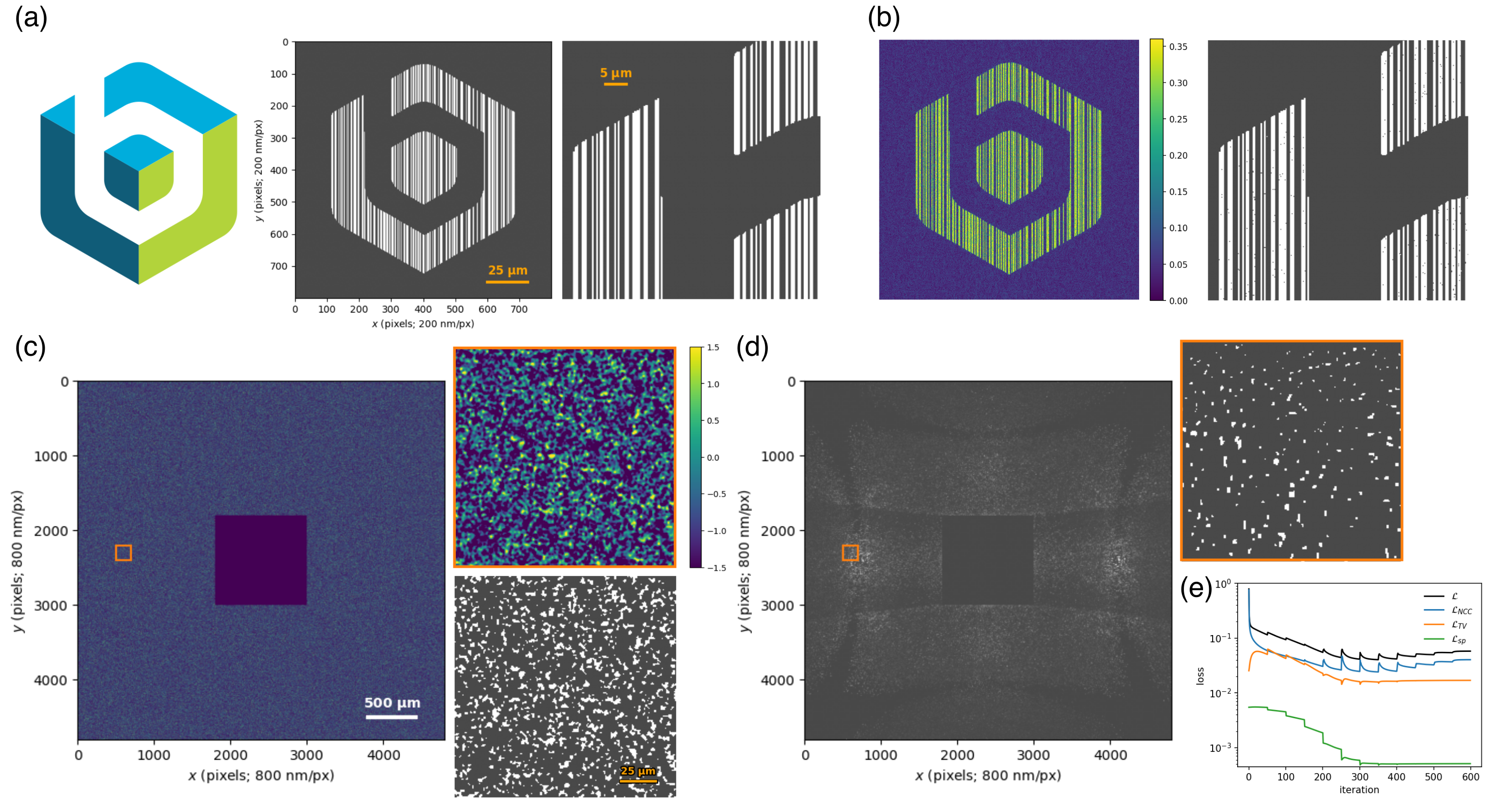}
   \end{center}
   \caption[example] 
   { \label{fig:asm_res} 
Mask design and simulated wavefront using ASM as the forward model. 
(a) The target pattern $I_t$ adapted from the Brookhaven National Laboratory (BNL) logo. Left: Original BNL logo. Middle: Binary target pattern consisting with stripes with varying width, with a minimum feature size of 400 nm (2 pixels). Right: zoomed-in view. 
(b) Left: simulated wavefront amplitude $A(x,y)$ produced by the optimized mask. Right: zoomed-in, binarized printed pattern $I_b$. 
(c) Initial logits $\ell(x,y)$ at the start of optimization, showing the central blocked region and the surrounding trainable region. 
The logits are defined on a coarse grid with pixel size 800 nm, and is upsampled before propagation. 
Top right inset: zoomed-in view of the orange square region. Bottom right: the binarized mask $M$. 
(d) Optimized mask $M$ after 600 iterations, with the inset showing over the same region as the insets in (c). 
(e) Iteration curve, showing the evolution of losses with iteration number. 
}
   \end{figure}

The target pattern $I_t$ is adapted from the Brookhaven National Laboratory (BNL) logo (Figure~\ref{fig:asm_res} (a)). 
It consists of vertical stripes of varying widths, with the thinnest line of 400 nm, corresponding to 2 pixels. 
To demonstrate a resolution gain from coherent diffraction, we define the mask logits on a coarser grid with pixel size $4\Delta x = 800$~nm. 
This ensures that the minimum feature size at the image plane is half that of the mask, 
The coarse mask is upsampled to the simulation resolution before propagation. 

\subsubsection{Mask Optimization}
\label{subsec:asm_opt}

The logits are initialized as a spatially correlated Gaussian random field with the central region fixed to be opaque throughout optimization, as described in Section~\ref{sec:optimization}. 
The initialized mask is shown in Figure~\ref{fig:asm_res} (c).
The mask logits $\ell(x,y)$ are optimized over 600 iterations using the Adam optimizer with a learning rate of 0.3. The total loss is 
$\mathcal{L} = \mathcal{L}_\mathrm{PCC} + w_1 \mathcal{L}_\mathrm{TV} + w_2 \mathcal{L}_\mathrm{sp}$, 
where $\mathcal{L}_\mathrm{TV}$ penalizes high-frequency spatial variation in the mask and $\mathcal{L}_\mathrm{sp}$ promotes mask sparsity. 
The regularization weights are $w_1 = 0.3$ and $w_2 = 0.02$.

The soft mask is given by $\tilde{M}(x,y) = \sigma\!\left(\ell(x,y) / \tau\right)$. 
To stabilize optimization, the sigmoid temperature parameter $\tau$ is annealed during training. 
We begin with $\tau = 2$ (soft mask) to allow stable gradient steps early in optimization, and reduce $\tau$ by a factor of 0.7 every 50 iterations, ending at $\tau = 0.04$ (nearly binary mask). 
This annealing schedule is reflected in the loss curve (Figure~\ref{fig:asm_res} (e)), which shows a small jump at every 50 iterations, since the harder sigmoid increases the loss before the optimizer recovers.

The optimization takes approximately 7 minutes on a single NVIDIA A100 GPU. The regularization terms drive the optimized mask to be sparser than the initialization, as visible in the zoomed-in view of the optimized mask (Figure~\ref{fig:asm_res} (d)).

\subsubsection{Performance}

Figure~\ref{fig:asm_res} (b) shows the propagated field amplitude $A(x,y)$ using the optimized mask. 
The Pearson correlation coefficient between $A$ and the target $I_t$ is 0.95. 
The stripe pattern is clearly resolved with high intensity contrast. 
The optimized mask with 800~nm pixel pitch produces features as small as 400~nm at the image plane, demonstrating a 2$\times$ resolution enhancement through coherent diffraction. 

In real experiments, resist image forms after the resist is illuminated by the propagated wave and after resist development. 
We approximate the resist response to light illumination by binarizing the intensity $I=A^2$ to generate the printed pattern: $I_b(x,y)$. The zoomed-in view of the binarized aerial image is shown on the right panel of Figure~\ref{fig:asm_res} (b). 
The binarized aerial image $I_b$ matches the target pattern well, with only isolated point defects accounting for $0.1\%$ of total pixels. The defects are randomly distributed. 

\subsection{Optimization Using Shifted ASM}

To resolve the issue of high memory requirements innate to optimizing with the ASM method, we employ the shifted ASM as the forward physics function. 
In this work, the mask is divided to  $ 4 \times 4$ patches for shifted ASM propagation.  
The optimization procedure does not change. 

\subsubsection{Comparison with ASM}

To verify that tiling gives equivalent propagation result under our conditions, we optimized a mask using the shifted ASM and then propagated the same mask with both ASM and shifted ASM. 
Figure~\ref{fig:shifted_asm} (a) shows the propagated field amplitude using ASM, and Figure~\ref{fig:shifted_asm} (b) 
shows the difference between the ASM and shifted ASM propagated amplitudes. 
The two are almost identical, with a Pearson correlation of 0.9998. 
This confirms that the shifted ASM and ASM are effectively equivalent under our padding ratio and patch count.



\subsubsection{Memory and Runtime Scaling}

Table~\ref{tab:comp-methods} compares the peak memory requirement and optimization runtime (600 iterations) using different methods. We use the NVIDIA A100-80GB GPU. 
The calculations were performed for a $19200\times19200$ mask size, with $30720^2$ pixels after padding. 
Using shifted ASM reduces the memory by $3.8\times$ for a $1.3\times$ increase in runtime. 
Adding gradient checkpointing further reduces peak memory to $7.4\times$ relative to ASM for $2\times$ the runtime. 
Since the working memory scales as $\mathcal{O}\big(N^2\big)$ for ASM and $\mathcal{O}\big((N/P)^2\big)$ for shifted ASM, the naive expectation is that using a $4\times4$ patch size will reduce memory usage to $1/16$. 
However, the logits and their backpropagation buffers, as well as the mask and its gradient, must all be stored in memory regardless of tiling. 
Therefore, the observed memory reduction falls short of the naive $1/16$ estimate.
The runtime barely increases, because the total area is conserved for ASM and shifted ASM, so the same total FFT work is performed. 
In fact, the FFT work for shifted ASM is marginally less, since the FFT runtime for $N^2$ pixels for ASM is $\mathcal{O}\big(N^2\log N^2\big)$ and for shifted ASM is $\mathcal{O}\big(N^2\log (N/P)^2\big)$. 
Overall, the results show that shifted ASM works with much smaller memory requirement without too much overhead time.    

\begin{table}[ht]
\caption{Working memory and optimization time from different methods. The unpadded mask size is $H=W=19200$. After padding, the size on each side is $N=30720$. } 
\label{tab:comp-methods}
\begin{center}       
\begin{tabular}{|l|l|l|}
\hline
\rule[-1ex]{0pt}{3.5ex}  Method & peak memory & wall time  \\
\hline
\rule[-1ex]{0pt}{3.5ex}  ASM & 46.84 GB & 409 s   \\
\hline
\rule[-1ex]{0pt}{3.5ex}  Shifted ASM, 1 GPU & 12.18 GB & 533 s  \\
\hline
\rule[-1ex]{0pt}{3.5ex}  Shifted ASM, 1 GPU, with checkpoint & 6.33 GB & 816 s  \\
\hline
\rule[-1ex]{0pt}{3.5ex}  Shifted ASM, 4 GPU & 7.00 GB / rank & 162 s  \\
\hline 
\end{tabular}
\end{center}
\end{table}

\begin{figure} [htbp]
   \begin{center}
   \includegraphics[width=0.8\textwidth]{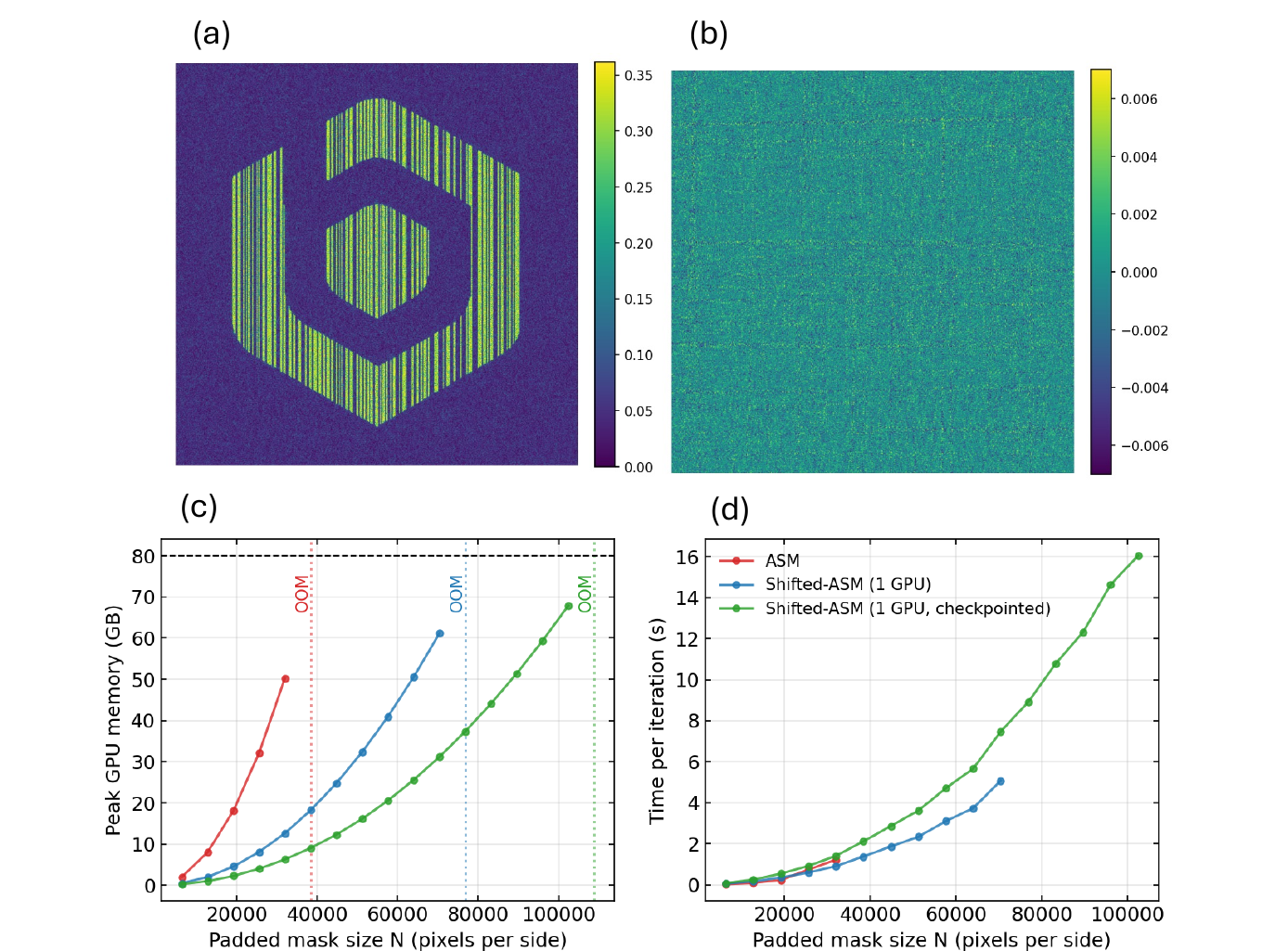}
   \end{center}
   \caption[example] 
   { \label{fig:shifted_asm} 
(a) Simulated field amplitude using the mask optimized with shifted ASM as the forward function, and propagated with ASM. 
(b) Difference between the ASM- and shifted-ASM-propagated field amplitudes. 
(c) Peak memory vs. mask size $N$. $N$ is the number of pixels per side after padding. 
(d) Time per iteration vs. padded size $N$. 
}
   \end{figure}

Figure~\ref{fig:shifted_asm} (c) and (d) show the peak memory and per-iteration time against padded size $N$ for ASM, shifted ASM, and checkpointed shifted ASM, each swept until the 80 GB GPU is out of memory (OOM).
The checkpointed shifted ASM reaches $N=102400$, corresponding to $H=W=64000$ before padding, a $12.8~\text{mm} \times 12.8~\text{mm}$ size mask. This is $10.2\times$ the mask area of the largest mask achievable with ASM.

\subsubsection{Multi-GPU Parallelization}
   
In shifted ASM, the complex wavefront of each sub-mask is summed up to produce the final pattern. 
The optimization can either be performed in serial using a single GPU, or in parallel using multiple GPUs, with patches distributed across GPUs. 
The only inter-GPU communication needed is two all-reduce operations: one summing the partial fields in the forward pass, and one summing the logit gradients after backpropagation. 
Distributing the 16 patches over 4 GPUs (without checkpointing), the optimization peak memory is 7.00 GB per GPU, and the optimization only takes 162 s (Table~\ref{tab:comp-methods}).

The current implementation does not change the peak memory using single or multiple GPUs when checkpointing is enabled. Because the peak memory is set by the mask plus the working memory for propagating and optimizing one patch, and distributing patches across GPUs does not reduce the memory usage. 
To further reduce the per-GPU memory, the mask can be sharded so that each rank stores only the $1/K$ of the mask with $K$ being the number of GPUs. 
We leave this to future work.

Together, these results demonstrate that shifted ASM reproduces ASM's output with negligible error. Using shifted ASM largely reduces peak memory with modest runtime cost, scales efficiently 
across multiple GPUs, and enables mask optimization at sizes inaccessible to standard ASM.

\section{Discussion}

\subsection{Monochromatic Beam}
\label{sec:bandwidth}

IL of periodic patterns is agnostic to wavelength~\cite{MOJARAD201555, Ekinci_Nanoscale_2024}, because the fringe period on the image plane is set by the pitch of the gratings on the mask and the mask-to-image distance. Hence periodic patterns can be printed with pink- or white-beam sources without loss of contrast. 
Nevertheless, this tolerance does not extend to arbitrary patterns, where the mask is irregular rather than having a single period. The printed pattern changes with wavelength, so a near-monochromatic beam is required for high-quality patterns. 

\begin{figure} [htbp]
   \begin{center}
   \includegraphics[width=0.96\textwidth]{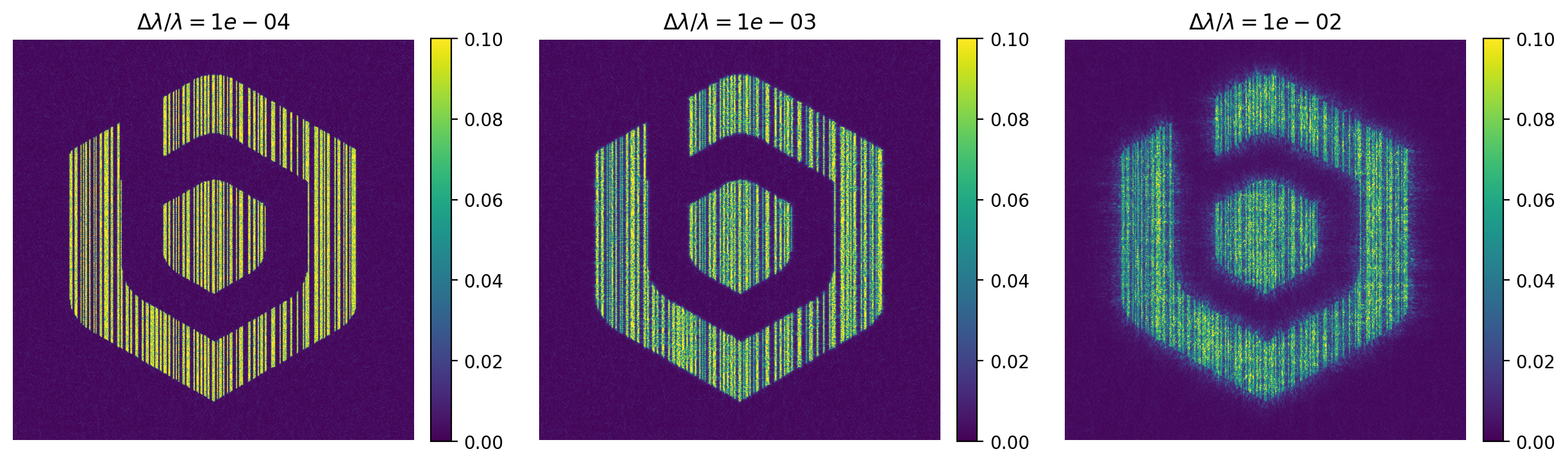}
   \end{center}
   \caption[example] 
   { \label{fig:bw} 
Propagated intensity using beams with different bandwidths. 
The correlation with the monochromatic result is 0.9999, 0.9243, and 0.8235 for $\Delta\lambda/\lambda = 10^{-4}$, $10^{-3}$, and $10^{-2}$, respectively. 
}
   \end{figure}

To quantify the sensitivity of bandwidth, we simulated the aerial image from an optimized mask under illumination with finite spectral bandwidth, using a Gaussian spectral distribution of mean wavelength $\lambda$ and standard deviation $\Delta\lambda$. 
Figure~\ref{fig:bw} shows the results at $\Delta\lambda/\lambda = 10^{-4}$, $10^{-3}$, and
$10^{-2}$. 
Here, we show the intensity rather than the amplitude, because the intensities at different wavelengths are summed incoherently to simulate the effects of finite bandwidth. 
At $10^{-4}$ the pattern is indistinguishable from the monochromatic design. At $10^{-3}$ the features begin to wash out, and at $10^{-2}$ the image is heavily
blurred. 

The threshold can be estimated from the coherence length of the beam, $L_c \approx \lambda^2/\Delta\lambda$. The optical-path difference (OPD) between beam diffracted from the mask center and from the mask edge is 0.77 mm for our geometry ($d = 2$~mm, $\mathrm{NA} = 0.69$). Requiring $\mathrm{OPD} \lesssim L_c$ gives $\Delta\lambda/\lambda \lesssim 5.76\times10^{-4}$, consistent with the simulation results. 
This requirement tightens with increasing NA and $d$. 
The monochromaticity of laser and synchrotron beam is well within the required range, but it rules out broadband sources acceptable for periodic IL.

\subsection{Mask Degeneracy}
\label{sec:mas-degen}

Because of the high degree of freedom of the mask, different mask designs can produce similar patterns that are close to the target. 
This degeneracy can be used to optimize toward masks with additional desirable properties. 

One application is to design masks that are more robust to imperfections in the experimental setup, including distance robustness and beam non-ideality robustness. 
For distance robustness, each iteration can sample $d$ from a range of distances close to the nominal value, and optimize the distance-averaged loss. 
The resulting mask trades image quality for tolerance to distance error, easing the sample-to-mask alignment in experiments. 
Nevertheless, the tolerance is limited as the finest feature size approaches the diffraction limit, where the depth of focus is small. 
Similarly, for robustness to beam non-idealities, each iteration can sample a beam profile with different fluctuations and optimize the average loss. This could produce masks that are less sensitive to beam variation between exposures, again at the cost to image quality. 

In addition, inverse problems are sensitive to initialization, so different initializations converge to different mask designs that produces similar patterns. 
For example, initializing the logits with fewer ``on'' pixels biases the optimization toward sparser masks, easing fabrication at the expense of reduced pattern intensity (i.e., less efficient use of the illumination).

\section{Conclusion}

We presented a gradient-based framework for inverse photomask design in IL for arbitrary, non-periodic target patterns, using the differentiable ASM as the forward propagator. 
The optimized mask reproduced a target pattern with a Pearson correlation of 0.95 and 0.1\% isolated pixel defects, resolving features half the mask resolution. 
To scale beyond the memory limits of standard ASM, we used a shifted-ASM-based optimization that reduces peak GPU memory by up to an order of magnitude with modest runtime overhead and scales efficiently across multiple GPUs. 
We further discussed two practical considerations, including the near-monochromatic illumination required to preserve pattern quality, and optimization strategies that could increase the robustness and manufacturability of the mask. 
We anticipate that this differentiable optimization framework could extend to other inverse problems wherever a forward differentiable model is available.
The source code is available at \url{https://github.com/chuntian236/holography-optimization.git}. 

\appendix

\section{Forward propagation models: ASM and shifted ASM}
\label{app:propagation}

This appendix gives the full formulation of the angular spectrum method (ASM) and shifted ASM propagators introduced in Sections~\ref{sec:ASM} and~\ref{sec:shifted-ASM}.

\subsection{Angular Spectrum Method}
\label{app:ASM}

Given the complex field amplitude $g(x,y;z=0)$ at the mask plane (source plane), ASM simulates the complex field $g(x,y;z=d)$ at the image plane (destination plane), with the following three steps:
\begin{enumerate}[label=\arabic*)]
    \item 
        Convert the source field $g(x, y; 0)$ to the angular spectrum $G(u, v;\, 0)$ using Fourier transform:  
        \begin{equation}
            G(u, v;\, 0) = \mathcal{F}\!\left\{g(x, y;\, 0)\right\}
            = \iint g(x,y;\,0)\, \exp\!\left[-i2\pi(u x + v y)\right]\, dx\, dy,
        \end{equation}
        where $u$ and $v$ are spatial frequencies. 

    \item 
        Multiply the angular spectrum by the transfer function to propagate to the image plane: 
        \begin{equation}
            G(u, v;\, d) = G(u, v;\, 0)\, H(u, v;\, d).
        \end{equation}
        The transfer function $H$ combines the propagation and the band-limiting window: 
        \begin{equation}
            H(u, v; d) = W(u, v) \exp\!\left[i 2\pi d 
            \sqrt{\lambda^{-2} - u^2 - v^2}\right].
        \end{equation} 
        The band-limiting window $W = 1$ inside allowed frequencies, and 0 outside. 
        For the simple case where only evanescent wave is filtered out, 
        $   W(u, v) = 1 $  for $ u^2 + v^2 \leq \lambda^{-2} $ and 0  \text{otherwise}. 

    \item 
        Recover the complex field at the image plane using inverse Fourier transform: 
        \begin{equation}
            g(x, y;\, d) = \mathcal{F}^{-1}\!\left\{G(u, v;\, d)\right\}
        \end{equation}        
\end{enumerate}

Besides the evanescent-wave filter, an aliasing-suppression filter is necessary due to finite sampling in simulation.   
When sampled on a discrete grid, the transfer function $H$ can cause aliasing because its local fringe frequency grows with propagation distance $d$ and can exceed the Nyquist limit set by the sampling interval. 
Following Matsushima~\cite{Matsushima:09}, for a source window with half-widths $S_x$, $S_y$, the bounds of the frequencies are: 

\begin{equation}
    u_{\mathrm{limit}} = \lambda^{-1}\!\left[1+\left(\frac{d}{S_x}\right)^{2}\right]^{-1/2},
    \qquad
    v_{\mathrm{limit}} = \lambda^{-1}\!\left[1+\left(\frac{d}{S_y}\right)^{2}\right]^{-1/2}.
    \label{eq:ulimit_centered}
\end{equation}

The anti-aliasing window becomes: 
\begin{equation}
    W(u,v) =
    \begin{cases}
        1 & \left(\dfrac{u}{u_{\mathrm{limit}}}\right)^{2} + (\lambda v)^{2} \le 1,\ \
            \left(\dfrac{v}{v_{\mathrm{limit}}}\right)^{2} + (\lambda u)^{2} \le 1,\ \
            u^2+v^2 \le \lambda^{-2} \\
        0 & \text{otherwise}.
    \end{cases}
    \label{eq:Wcentered}
\end{equation}

A larger propagation distance $d$ relative to the padded window size tightens this 
bound, clipping more of the spectrum to avoid aliasing.

\subsection{Shifted ASM}
\label{app:shifted-ASM}

For a destination window shifted by ($x_0, y_0$) from the optical axis, the modified transfer function is: 
\begin{equation}
    \hat{H}(u, v;\, d) = \hat{W}(u,v)\exp\!\left[i2\pi\left(x_0 u + y_0 v + d\sqrt{\lambda^{-2} - u^2 - v^2}\right)\right], 
\end{equation}
where $\hat{W}$ is now the shifted band-limiting window and depends on the offset ($x_0, y_0$) and propagation distance $d$~\cite{Matsushima:10}.

Following Matsushima~\cite{Matsushima:10}, the bounds to limit the frequencies are: 
\begin{equation}
    u_{\mathrm{limit}}^{(\pm)} = \lambda^{-1}\!\left[1+\left(\frac{d}{x_0 \pm S_x}\right)^{2}\right]^{-1/2},
    \qquad
    v_{\mathrm{limit}}^{(\pm)} = \lambda^{-1}\!\left[1+\left(\frac{d}{y_0 \pm S_y}\right)^{2}\right]^{-1/2},
    \label{eq:ulimit_shifted}
\end{equation}
which reduce to Eq.~\eqref{eq:ulimit_centered} when $x_0=y_0=0$.

The anti-aliasing band-limiting condition on $u$ becomes:
\begin{equation}
    \begin{cases}
        u \ge 0,\ \
        \left(\dfrac{u}{u_{\mathrm{limit}}^{(-)}}\right)^{2} + (\lambda v)^{2} \ge 1,\ \
        \left(\dfrac{u}{u_{\mathrm{limit}}^{(+)}}\right)^{2} + (\lambda v)^{2} \le 1
        & S_x < x_0, \\[3mm]
        \left(\dfrac{u}{u_{\mathrm{limit}}^{(-)}}\right)^{2} + (\lambda v)^{2} \le 1 \ \text{for } u \le 0, \quad \text{or} \quad
        \left(\dfrac{u}{u_{\mathrm{limit}}^{(+)}}\right)^{2} + (\lambda v)^{2} \le 1 \ \text{for } u > 0
        & -S_x \le x_0 < S_x, \\[3mm]
        u \le 0,\ \
        \left(\dfrac{u}{u_{\mathrm{limit}}^{(+)}}\right)^{2} + (\lambda v)^{2} \ge 1,\ \
        \left(\dfrac{u}{u_{\mathrm{limit}}^{(-)}}\right)^{2} + (\lambda v)^{2} \le 1
        & x_0 \le -S_x.
    \end{cases}
    \label{eq:ucondition_shifted}
\end{equation}
The condition on $v$ is symmetric to that of $u$, and can be obtained by changing $u \to v$, $x_0 \to y_0$, $S_x \to S_y$, and $u_{\mathrm{limit}}^{(\pm)} \to v_{\mathrm{limit}}^{(\pm)}$. 

The resulting band-limiting window $\hat{W}(u,v)$ is narrower and off-centered relative to 
$W(u,v)$ in Eq.~\eqref{eq:Wcentered}, in proportion to how far the offset $(x_0,y_0)$ moves the destination window from the optical axis.

\section{Mask optimization loss function}
\label{app:loss}

At each optimization iteration, the propagated wavefront is calculated using ASM or shifted ASM, and the simulated field amplitude $A(x,y) = |g(x,y;\,d)|$ is compared with the binary target pattern $I_t(x,y)$ using the total loss: 
\begin{equation}
    \mathcal{L} = \mathcal{L}_\mathrm{PCC} + w_1\, \mathcal{L}_\mathrm{TV} + w_2\, \mathcal{L}_\mathrm{sp},
\end{equation}
where the Pearson correlation loss is: 
\begin{equation}
    \mathcal{L}_\mathrm{PCC} = 1 - 
    \frac{\displaystyle\sum_{x,y} \left[A(x,y) - \bar{A}\right]\left[I_t(x,y) - \bar{I}_t\right]}
    {\sqrt{\displaystyle\sum_{x,y}\left[A(x,y) - \bar{A}\right]^2 \cdot \sum_{x,y}\left[I_t(x,y) - \bar{I}_t\right]^2}},
\end{equation}
the total variation regularization is: 
\begin{equation}
    \mathcal{L}_\mathrm{TV} = \frac{1}{N_M}\sum_{x,y} \left[ 
    \left|\tilde{M}(x+1,y) - \tilde{M}(x,y)\right| 
    + \left|\tilde{M}(x,y+1) - \tilde{M}(x,y)\right| 
    \right],
\end{equation}
and the sparsity regularization is: 
\begin{equation}
    \mathcal{L}_\mathrm{sp} = \frac{1}{N_M}\sum_{x,y} \tilde{M}(x,y).
\end{equation}

Here, $\bar{A}$ and $\bar{I}_t$ are the spatial average of $A$ and $I_t$, respectively, $N_M$ is the total number of mask pixels, and $w_1$ 
and $w_2$ are scalar weights of the regularization term. Section~\ref{subsec:asm_opt} reports the values of $w_1$ and $w_2$ used in this work.
$\mathcal{L}_\mathrm{TV}$ penalizes high-frequency spatial variation in the mask and $\mathcal{L}_\mathrm{sp}$ promotes mask sparsity, both favoring masks that are easier to fabricate.

\acknowledgments 

This research was supported by Brookhaven National Laboratory (BNL), Laboratory Directed Research and Development (LDRD) Grant No. 25-038, ``Co-Design of Advanced Interference EUV Lithography Capability for Next Generation Semiconductor Chip Manufacturing''.
This research was partially funded by BNL LDRD Grant No. 24-057, ``Development of Holistic and Scalable Solutions to Microelectronics Metrology
Challenges''.
This research used resources of the Scientific Computing and Data Facilities (SCDF) at BNL. 
We thank Dr. Longlong Wu for valuable discussions. 
We acknowledge the use of Claude Code (Anthropic) to assist with debugging and refining of the mask optimization code. All code and results were verified by the authors, who take full responsibility for the content.

\bibliography{ref} 
\bibliographystyle{spiebib} 

\end{document}